\documentclass{ws-ijmpc}
\usepackage[super]{cite} % other options `nocompress`, `nosort`

\begin{document}

\title{Spatial interference between infectious hotspots: epidemic condensation and optimal windspeed}
\author{Johannes Dieplinger}
\address{Scuola Normale Superiore di Pisa, Piazza dei Cavalieri, Pisa, Italy\\
johannes.dieplinger@t-online.de}
\author{Sauro Succi}
\address{Center for Life Nanosciences at La Sapienza,  Italian Institute of Technology, Roma, Italy,\\
Scuola Normale Superiore di Pisa, Piazza dei Cavalieri, Pisa, Italy,\\
Physics Department, Harvard University, Cambridge, USA,\\
sauro.succi@sns.it}

\title{Spatial interference between infectious hotspots: epidemic condensation and optimal windspeed}

\maketitle

\begin{abstract}
	We discuss the effects of spatial interference between two infectious hotspots as a function
	of the mobility of individuals (wind speed) between the two and their relative degree of infectivity.
	As long as the upstream hotspot is less contagious than the downstream one, increasing the
	wind speed leads to a monotonic decrease of the infection peak in the downstream hotspot.
	Once the upstream hotspot becomes about between twice and five times more infectious than the 
	downstream one, an optimal wind speed emerges, whereby a local minimum peak intensity is 
	attained in the downstream hotspot, along with a local maximum beyond which 
	the beneficial effect of the wind is restored.  
	Since this non-monotonic trend is reminiscent of the equation of state of non-ideal
	fluids, we dub the above phenomena "epidemic condensation".
	When the relative infectivity of the upstream hotspot exceeds about a factor five, the beneficial effect
	of the wind above the optimal speed is completely lost: any wind speed above the optimal one
	leads to a higher infection peak. It is also found that spatial correlation between the
	two hotspots decay much more slowly than their inverse distance.
	It is hoped that the above findings may offer a qualitative clue for optimal confinement
	policies between different cities and urban agglomerates.  
	\keywords{population dynamics, SIR model, interference, nonlinearity, pattern formation}
\end{abstract}

\ccode{PACS Nos.: 87.23.Cc, 01.75.+m, 02.70.Bf}

\section{Introduction}

In early 2020 the world has been taken by a very aggressive global 
pandemic, the covid-19, which spread around the entire planet 
at unprecedented speed in mankind history.
As we speak, the pandemic, originated in Wuhan, China, allegedly in January 2020
has spread out over 100 countries, with over ten million contagion cases and over 500,000 casualties 
worldwide, as of end June 2020 \cite{covidnumbers}, giving rise to the trade-off between strong confinement measures to 
stave off devastating effects on health systems 
\cite{Armocida2020,Legido-Quigley2020} and economic, social and psychological terms \cite{Chen2020a, Duan2020, Nicola2020, Toda2020}.

A distinctive feature of the covid-19 pandemia is the heterogeneity of
the viral infection in space; in many countries a large fraction of the overall infection counts 
originated from very specific hotspots, such as Lombardy in Italy and NYC in the USA.
This strong inhomogeneity calls, among others, for a proper modelling of the mechanism 
by which the infection propagates in space and time \cite{Lloyd1996}.  

In this paper we address this issue by isolating a toy-problem, namely the 
interaction between two infectious hotspots sitting at two separate locations in space. 
Special attention is paid to the spatial interference \cite{Dietz1979} between the two 
hotspots, in particular the way that the presence of the first affects the viral evolution 
in the second, depending on the mobility and infection rates in the two hotspots.

To this purpose, spatial mobility is described by a simple advection-diffusion SIR (ADSIR) model, in 
which diffusion encodes small-distance mobility (say walking), while advection stands 
for mid-range mobility (say train or car driving). { Rather than being an exact model of these real-world mobility schemes -- for which network models would certainly be a more accurate choice -- diffusion and advection have the general purpose to realize two different mobility mechanisms with a different scaling in time and space. }
Even though human mobility proceeds by more complex mechanisms than AD, typically encoded
by mobility networks \cite{Ball2010,Lang2018,VespiCovid1,VespiCovid2},  the present ADSIR model exposes nonetheless 
a number of interesting qualitative features related the spatio-temporal coupling between the two infectious hotspots.

\section{Mathematical formulation} We describe the covid-19 pandemic by means of a standard SIR model 
\cite{Kermack1927} -- which is the starting point for numerous interesting models in epidemiology  \cite{Kaushal2020,Kaxiras2020,Pathak2010,Ji2011,Lang2018,Roda2020, Toda2020,Calafiore2020} -- coupled in space via an advection-diffusion mechanism: 

\begin{eqnarray}
\label{ADSIR}
\partial_t s  = \nabla (-U s + D \nabla s)   -\beta si\\
\partial_t i  =  \nabla (-U i  + D \nabla i)   +\beta si -\gamma i\\
\partial_t r  = \nabla (-U r  + D \nabla r)  +\gamma i\  
\end{eqnarray} 
where $sir(x,y,t)$ is the population of Susceptible, Infected and Recovered individuals
at position $x,y$ and time $t$, respectively.
The coefficients $\beta,\gamma$ correspond to infection and 
recovery rate, respectively. 
In the above equations $U$ is the wind speed, which we take aligned with the x-axis
without loss of generality and $D$ is the diffusivity. 
The total number of individuals of species $k=s,i,r$ at a given time $t$ is thus given
by integral over the entire domain of the corresponding densities:
$
N_{k,h}(t) =  \int_{H_i} n_k(x,y,t) dxdy, \;k=s,i,r, \;i=1,2
$
where the integral runs over the hotspot regions HS1 and HS2.
% ---------------------------------   Fig 1 ------------------------------
\begin{figure}[t!]
	\centering
	\includegraphics[width=\linewidth]{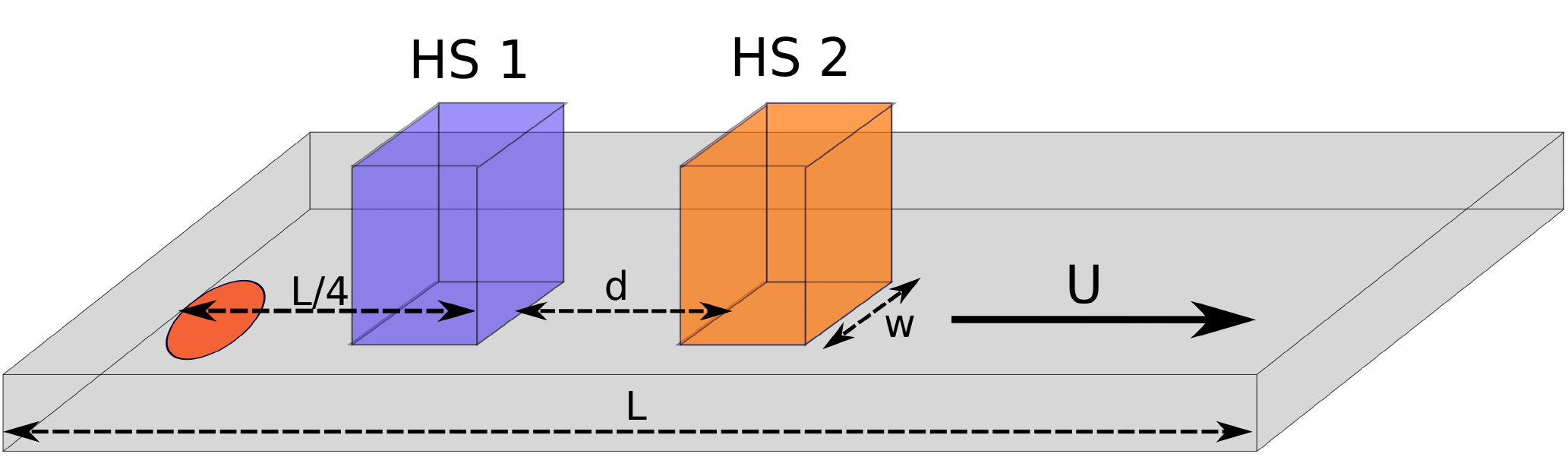}
	\caption{{The simulation domain.} Schematically shown is the simulation domain of length $L$. 
	The boxes indicate the hotspots HS1 and HS2 with different contagion rate $\beta_i=f_i\times \beta_0$, where $\beta_0$ is the 
	basic contagion rate in the normal domain. 
	The hotspots at distance $d$ have width $w$ in both directions. 
	The homogeneous wind $U$ is indicated by the black solid arrow and the red area indicates the small outbreak. }
	\label{fig:domain}
\end{figure}
The speed $U$ will be compared to an important intrinsic reference velocity:\\

The infected population $i(t)$ grows in HS1 as
$i(t)\approx A\cdot e^{(s(t)/N\cdot f_1\cdot\beta_0-\gamma)t}$.
Whenever the wind exceeds a critical speed $U_c=w\cdot (s(t)/N\cdot f_1\cdot\beta-\gamma)$ 
mitigation of epidemics is expected, due to the removal of infected individuals from the hotspot. 

For $U/U_c>1$, we expect the first hot-spot to become transparent and have little influence on the second one.
This of course largely depends on the a priori unknown parameters $s(t)/N$ and the variable parameter $f_1$. 
For a reference, we take $f_1=f_2=50$ and $s(t)/N\approx0.5$.

Hence, we define the reference speed $U_r=w\cdot(\frac{f_2\beta}{2}-\gamma)$.

The main independent (dimensionless) parameters are then defined as follows:
The contagion rate is $\beta_0=0.2$ in the normal domain and $\beta_i=f_i\cdot \beta_0$ in the hotspots $i=1,2$.
The recovery rate is homogeneous $\gamma=0.15$, such that the reproduction factor 
is $R_i=f_i\cdot \frac{\beta_0}{\gamma}>1$.
We fix $f_2=50$ while $f_1$ ranges from $1$ to $500$, so that the relative 
infectivity ratio $f=f_1/f_2\in\left(0.02,10\right)$.
The diffusion coefficient is $D=5$.
The width of the hotspots is $w=10$ and their distance is $d=50$.
The wind speed is measured in units of the reference speed, i.e. $u=U/U_r$. 
The grid spacing is $1\ \text{km}$ and time is measured in days, corresponding to a reference speed 
$U_r=50\ \text{km}/\text{day}$  and a diffusivity $D=5\ \text{km}^2/\text{day}$. These are plausible scales for human mobility. 

%We distinguish the following parameter regimes.
%Wind speed: no wind (strict lockdown, $u=0$), mild (mild lockdown $u<1$), strong (no lockdown $u>1$).
%HS2 infectivity:  weak ($f \ll1$), mild ($f<1$),  strong  ($f>1$).

We then study the solution of the ADSIR problem above as a function of the parameters $u$ and $f$.

In particular, we wish to assess under what conditions the presence of HS1 causes an increase
of infections in HS2 in terms of both peak intensity and duration.

\section{Simulation setup and results} We set up two hotspots HS1 and HS2 of width $w$ at position $x_1=L/4$ and 
$x_2=x_1+d+w$ respectively. 
The domain is a grid of size $64\times 1024$. 
We place a small Gaussian outbreak at $x=0$ and $y=W/2$ with a cutoff at $x=10$ in 
order to ensure that there are no initial infections in the hot-spots. 
The boundary conditions are chosen to be fully periodic. 
Fig. \ref{fig:domain} shows the geometric set-up of the hot-spots in the domain.

%{Definition of epidemic peak and duration}

\begin{figure}[t!]
	\includegraphics[width=\linewidth]{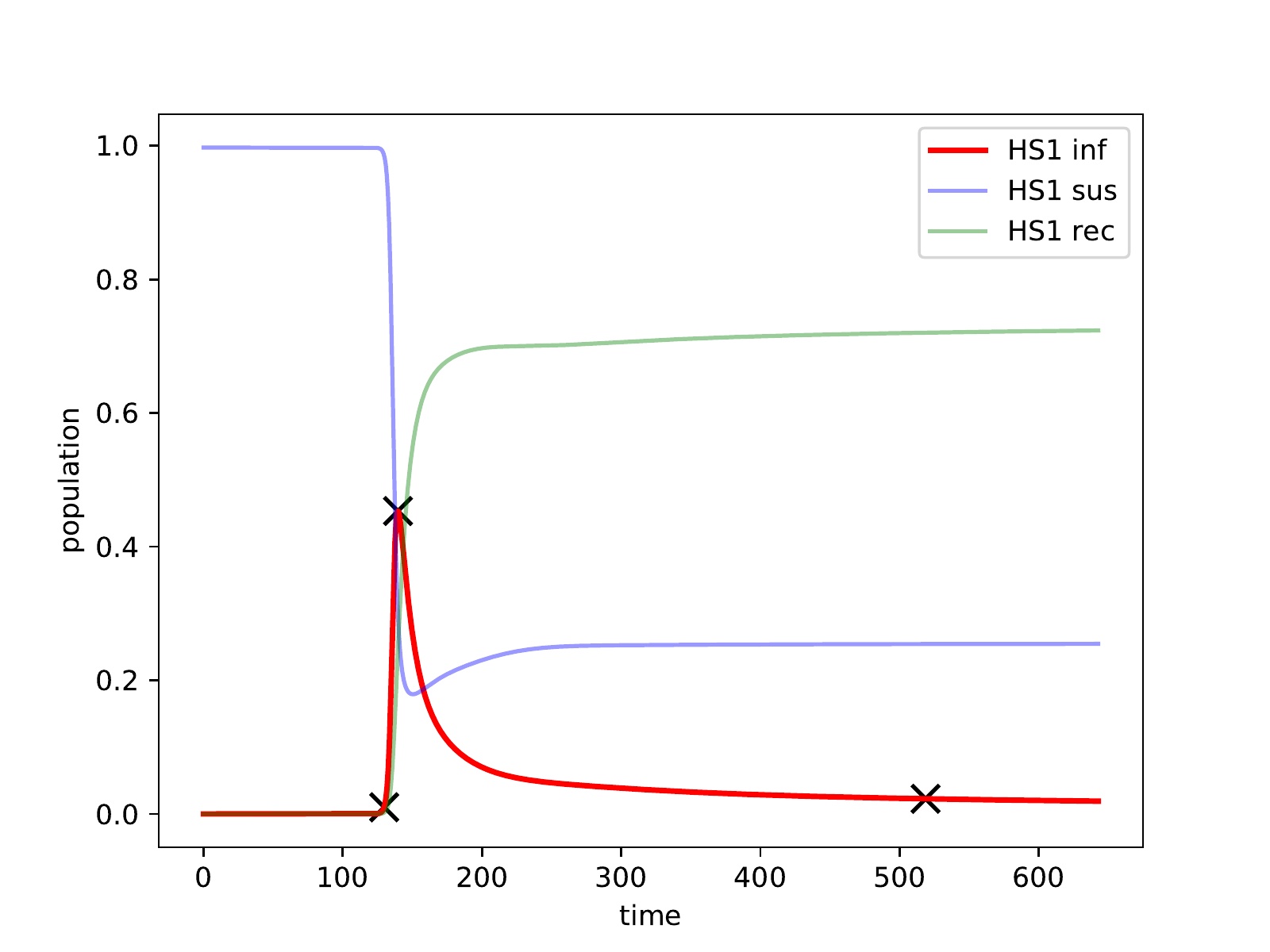}
	\caption{{Time evolution of the SIR populations in a single hotspot.} 
		The baseline parameters are: $f_1=f_2=5$, $\beta=0.5$, $U=0$, $D=0.05$. 
		The position of the hot-spot is at $x=L/4$. 
		The red curve shows the infected population in the hotspot, the black x's mark the start-up time, peak time and the decay time of the local epidemic, respectively. }
	\label{fig:times}
\end{figure}

In Fig. \ref{fig:times} we show the typical evolution of the infected species in a single hotspot.
The starting time of the hot-spot is defined as the time $t_s$ at which the 
infected population reaches $i(t_s)=0.002\cdot N$ and the end time $t_e$ at which the 
infected population has dropped to $i(t_e)=0.05\cdot i_{max}$. 
The time difference $\Delta t=t_e-t_s$ can be defined as the duration of the epidemic in this region. \\

%The peak intensity $I_{max}$ is a primary concern for short-term impact on the health system, whereas
%the duration $\Delta t$ has a bearing on the mid-long term impact on the economic-social system.

% ------------------- Fig 2: typical time evolution

%\subsection{Spatial interference}

\begin{figure}[t!]
	\centering
	\includegraphics[width=\linewidth]{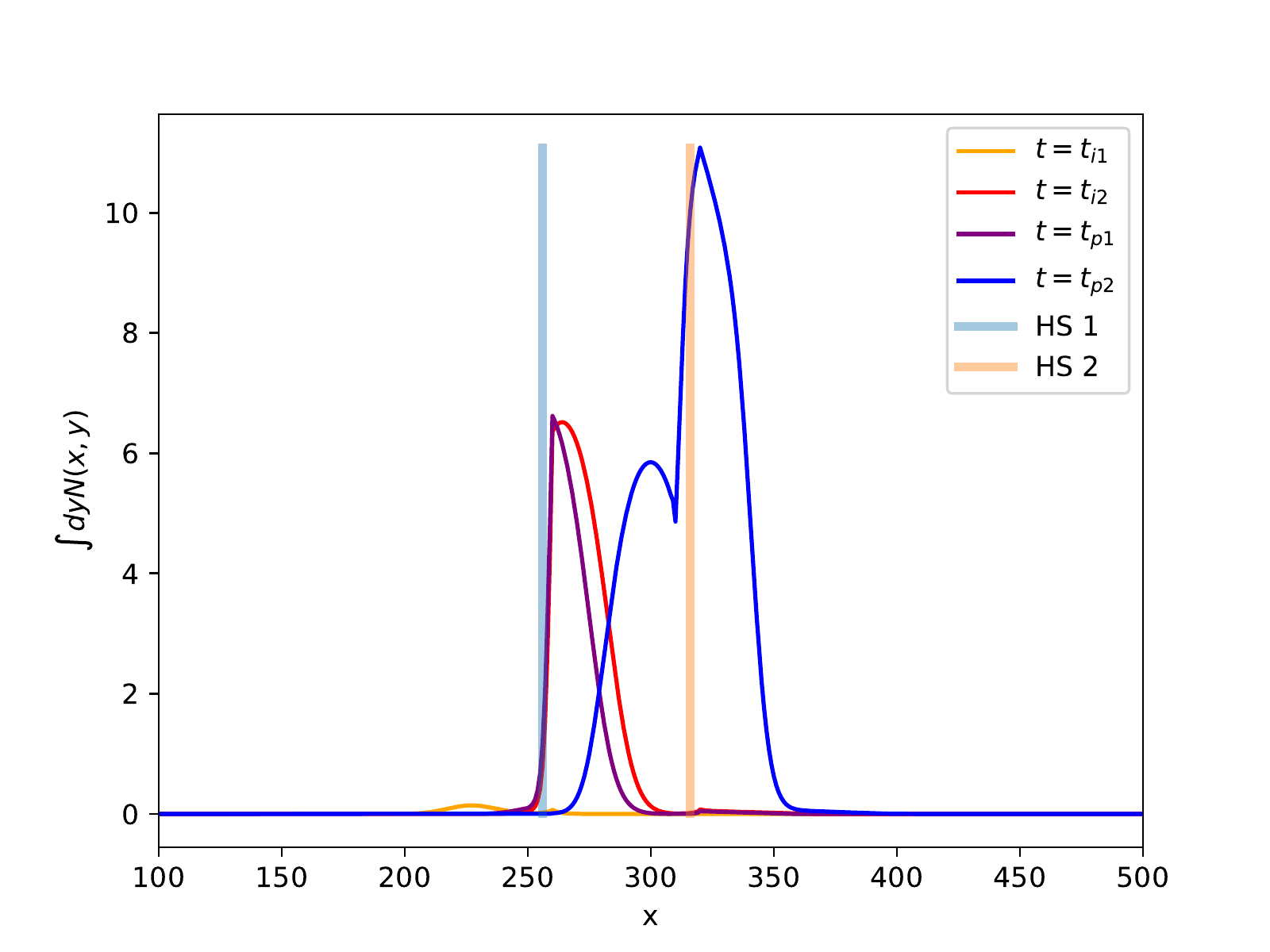}
	\caption{{Spatial distribution of the infected population.} 
		The y-integrated population with respect to the $x$-position. 
		The hotspots are marked with vertical lines. 
		The parameters are $f_1=f_2=50$, $\beta=0.2$, $\gamma=0.15$, and $u=0.28$, $d=50$, $w=10$, $D=0.05$.
		The snapshots are taken at the start-up time of both hotspots and the peak time of the second one. 	
		The development of a spatial interference between the two hotspots is clearly visible.
		The HS2 peak for the homogeneous case is  $0.99 \times 10 = 9.9$, against an observed one 
		of about 12, showing a 20 percent increase due to spatial interference from HS2.}
	\label{fig:profile}
\end{figure}

In Figure \ref{fig:profile}, we report a typical spatial interference pattern between HS1 on HS2.
The figure clearly shows that the infected population generated in HS1 reaches up to HS2 
and increases the local infection rate, thereby increasing the peak and possibly the duration as well. 

This is the typical scenario that HS2 policy makers endeavour to combat via lock-down measures.

% -----------------  Fig 3 ------------------ Spatial profiles

\begin{figure}[t!]
	\includegraphics[width=\linewidth]{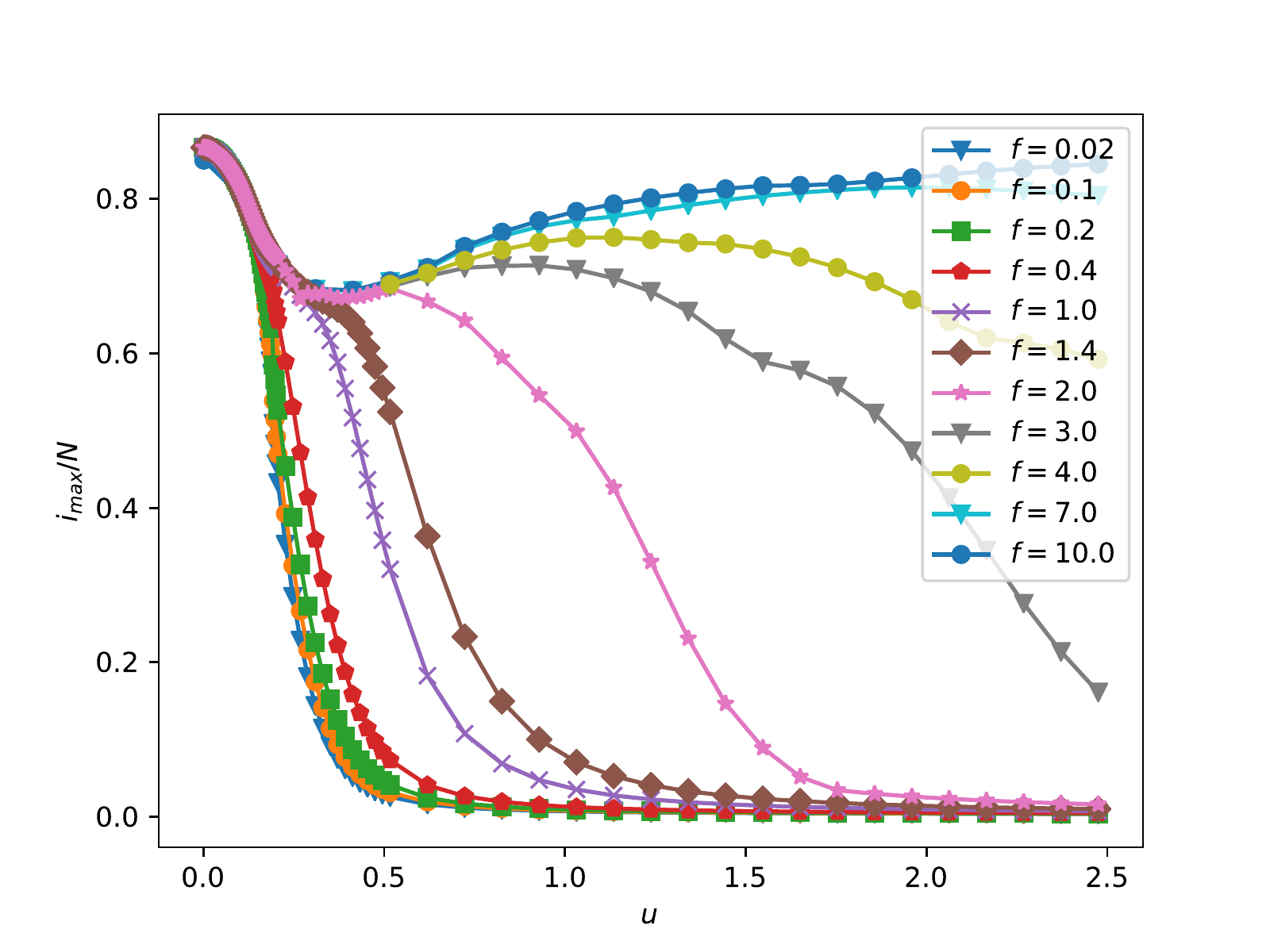}
	\caption{{Peak intensity in HS1 as a function of the wind speed at varying the infection rate in HS2} 
		The simulation parameters are the same as in figure	3.
		We clearly observe the emergence of a non-monotonic wind speed regime in the range $1<f<5$,
                 followed by a loss of any beneficial wind effect above $f \sim 5$.
	}
	\label{fig:close_wind}
\end{figure}

\textit{Peak and duration of the epidemics.}--In Fig. \ref{fig:close_wind}, we summarize the effect of the wind and HS1
infectivity on the peak intensity of HS2.

% --------------------------------------- Fig 4: main result
% -------------------------------------------------------------
%\textit{Shadow effect.}--

We measure the dimensionless peak value $i_{max}/N$ in the second (downstream) hotspot 
as a function of $u$ and $f$, where $N$ is the total number of individuals in the hotspot and $i_{max}$ 
is the peak value of the infected population. \\\

A few comments are in order.

First, we see that the peak intensity is a fast decreasing function of the wind speed for all HS1 infection ratios
well below the HS2 values. This is expected, since the infected in HS2 get replaced by less infected from HS1.

However, upon increasing $f_1$ in the vicinity and then above $f_2$, a shoulder 
appears at intermediate wind speeds, indicating that a highly contagious mobile population from
HS1 is capable of spoiling the beneficial effect of the wind. 
This is also a plausible result, since the infected removed by the wind 
in HS2 are quickly replaced by even more infected transmitted by HS1. 
This is the typical scenario dreaded by southern Italy towards the "stampede"
from northern regions in the early stage of the Italian epidemics.

Our simple model shows that such fears were indeed justified, an infectivity 
ratio $f=2$ is already capable of producing a secondary peak in the curve and
raising $f$ only makes the situation worst, with the emergence of a whole 
range of wind speeds in which the peak intensity grows instead of decaying,
almost reaching up to the value of the windless case.

Since this strongly reminds of the unstable region of a non-ideal equation 
of state, in which pressure goes down upon increasing density 
(condensation), we dub this  effect  "epidemic condensation".

This is the main result of this paper, as it highlights the existence of an 
optimal wind speed $u_{min} \sim 0.5$ which minimises the HS2 peak, and a second, higher,
characteristic speed $u_{max}$, beyond which the beneficial effects of the wind are restored. 
By further increasing the relative infectivity of HS1, between five and ten, no decay of the peak 
intensity at increasing wind speed above $u_{min}$ is observed anymore {in the simulated window of the wind speed $u$}, indicating that 
the presence of HS1 completely cancels any benefit of the wind speed above the optimal value $u_{min}$. {However, for a very large wind speed $u\to\infty$ we expect again a decrease of the infected ratio, since for infinite wind speed the hot-spots become transparent again, and infectivity will have no impact.}
  
%Once the size and the infection rate of the HS2 are known, one immediately identifies
%the optimal wind speed speeds in which the wind looses its beneficial effect. 

%\subsection{Duration of the epidemics}

\begin{figure}[t!]
	\includegraphics[width=\linewidth]{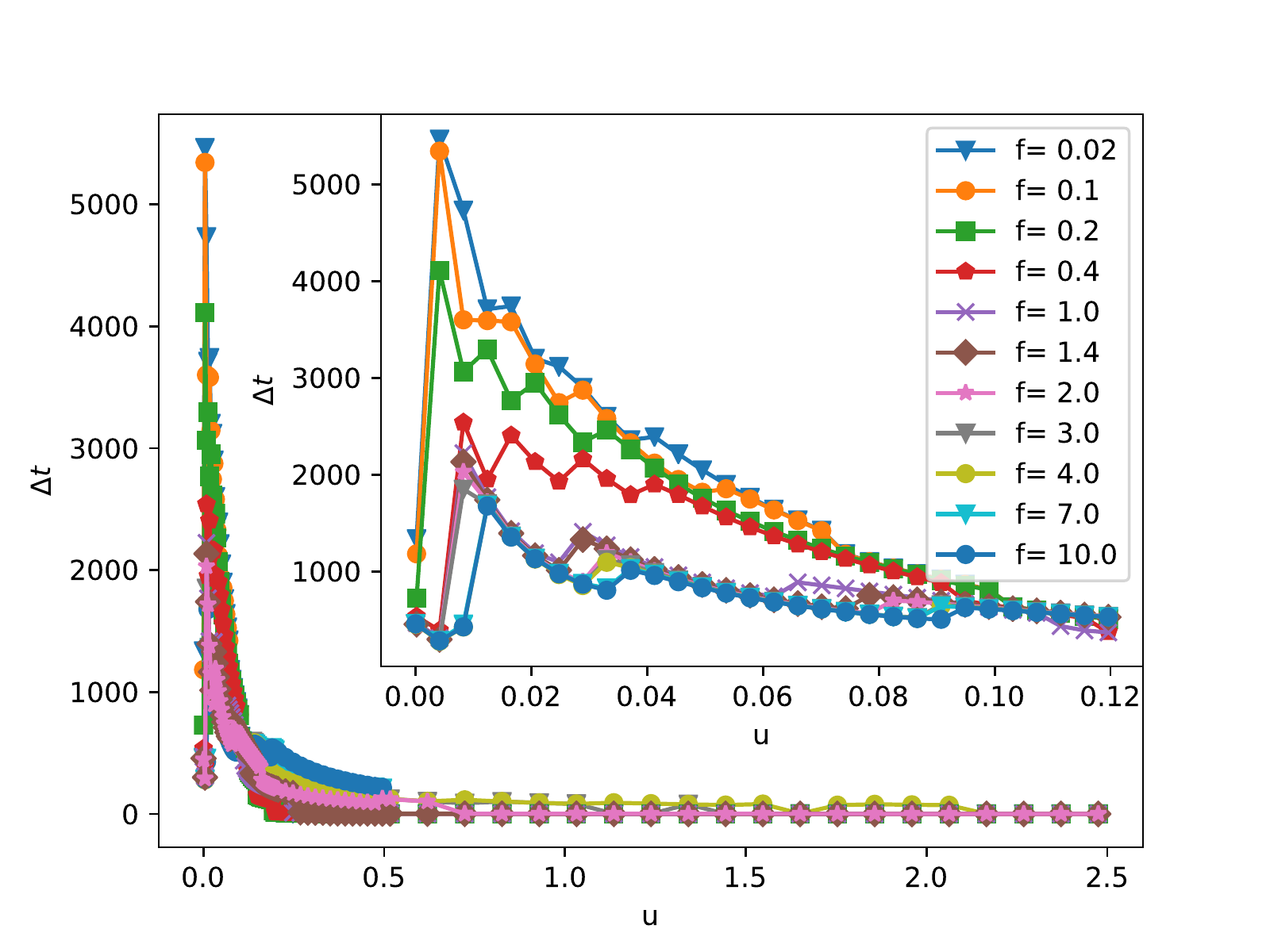}
	\caption{{Duration of the epidemic in HS2.}  
		The curve shows a peak at very low wind speeds 
		$u \sim 0.01$, followed by a sequence of secondary peaks
		at higher speeds, all well below $u_{min}$.
		By and large,  wind speeds above $u=0.1$ are consistently beneficial in shortening the epidemic duration.
	}
	\label{fig:dur_u}
\end{figure}

In Fig. \ref{fig:dur_u} we report the duration of the epidemics as a function of $u$ and $f$.
A major peak is observed at low-wind, corresponding to the fact that the 
infected are convected away at very low rates. As expected, high 
infectivity goes with high peaks and short durations, the dreaded
scenario for intensive care departments.  

As the wind speed increases, the local infected are efficiently removed
and the epidemic duration shortens. However, starting from comparatively low infectivity ratios,
$f=0.2$, further satellite peaks appear, indicating the existence of a sequence
of wind speeds such that the duration grows back, if only mildly. 
This is again interpreted as a spatial interference effect, although we must 
caution that such measurement is very sensitive to small changes of the 
duration threshold, hence should be taken with great caution.

% --------------------------------   Fig 5
% -----------------------------------------------------------

%Results of figure \ref{fig:dur_u}:
%\begin{itemize}
%	\item There seems to be an optimal wind speed at which the duration is the longest
%	\item Strange discontiniuties: return of the virus
%	\item the critical speed seems not to be related strongly to $u=1$, there seems to be a second also important critical wind speed $U_{c2}$
%	\item Explanation ?
%	
%\end{itemize}
% ------------------------------------------------------------------
% --Appendix: Eloborating the effect of the inter-hotspot distance--
\section{Qualitative scenario and discussion} The ADSIR model presented in this paper focusses on the effects of spatial 
coupling, advection and diffusion, on epidemic growth as dictated by local infection rates. 
It is well known that in the presence of random heterogeneities, such coupling 
can lead to highly nontrivial behaviour, such as the formation of striated infection highways \cite{Chotibut2017}.

Here we take a simpler model problem, namely the effect of a primary hotspot (HS1) on the epidemic growth 
on a secondary hotspot (HS2) downstream HS1. In particular, we focus on the effect of a uniform "wind" 
at speed $u$, mimicking a uniform human mobility across the two hotspots.

In the absence of any wind, $u=0$, and discounting diffusion, the two hotspots
evolve independently based on their corresponding infection rates.

As soon as the wind is switched on, a beneficial effect is 
expected for both HSs because the wind sweeps infected individuals away into
the "country side", where the chance to infect is much lower and healing can proceed nearly undisturbed.
This is certainly true as soon as the wind speed exceeds the infection speed,
namely the size of the hotspot divided by the typical infection timescale 
(reference speed), because, under such conditions, the wind blows 
susceptible individuals away before they have time to get significantly infected.

So, the baseline expectation is that  "wind is good", as it gives no
time for infection to develop substantially.
This is true for HS1, but not necessarily for HS2, which is 
exposed to the incoming flux of infected individuals from HS1. 

The quantitative question is whether, from the HS2 perspective, there exists
an optimal  wind speed which corresponds to a local minimum of the infection peak . 

In the following, we shall present evidence that the answer  
is in the affirmative. In particular, it is shown that as soon as
HS1 is more infectious than HS2, the peak intensity in HS2 develops 
a much slower decay with the wind speed, and when HS1 is significantly 
more infectious than HS2, the HS2 peak increases at increasing
wind speed, before it starts to decay again in the strong wind regime.
In other words, the HS2 peak develops a non-monotonic dependence on the 
wind speed, with a local minimum, $u_{min}$ at about half the reference speed
and a local maximum $u_{max}$ about twice as large.

Such non-monotonic dependence bears an intriguing resemblance
to a non-ideal equation of state, with the  unstable branch in the wind speed
region $u_{min} \le u \le u_{max}$.
Because of this close resemblance to equation of state of non-ideal gas, 
and most notably to the unstable region where a density increase leads to
a pressure decrease (condensation), we dub this effect {\it epidemic condensation}.

We also monitor the duration of the epidemics as a function of the wind speed and infection rates.
Note that while the peak intensity is the prime concern for health
capacity issues, the duration bears directly on the mid-long term
policies towards social and economic impact (many countries insisted
on "curve flattening" policies).  

Again, we find that wind increase above a very low threshold is generally 
beneficial, although  at increasing HS2 infectivity, the duration increases and shows 
repeated small-amplitude "sawtooth" oscillations. 
Such oscillations are yet another signature of spatial coupling, although their specific
nature remains to be fully ascertained.

\section{Effect of the hotspot distance and the diffusivity} 
We also inspected the effect of the hotspot distance on epidemic condensation.  
To this purpose, we ran a series of simulations at different wind speeds and  
distances in the range $50 \le d \le 200$, keeping a fixed value $f=4$.
 
 As expected, the local  maximum observed in the condensation decreases with the distance and, less expectedly, so does the local minimum. 
\begin{figure}[t!]
	\centering
	\includegraphics[width=\linewidth]{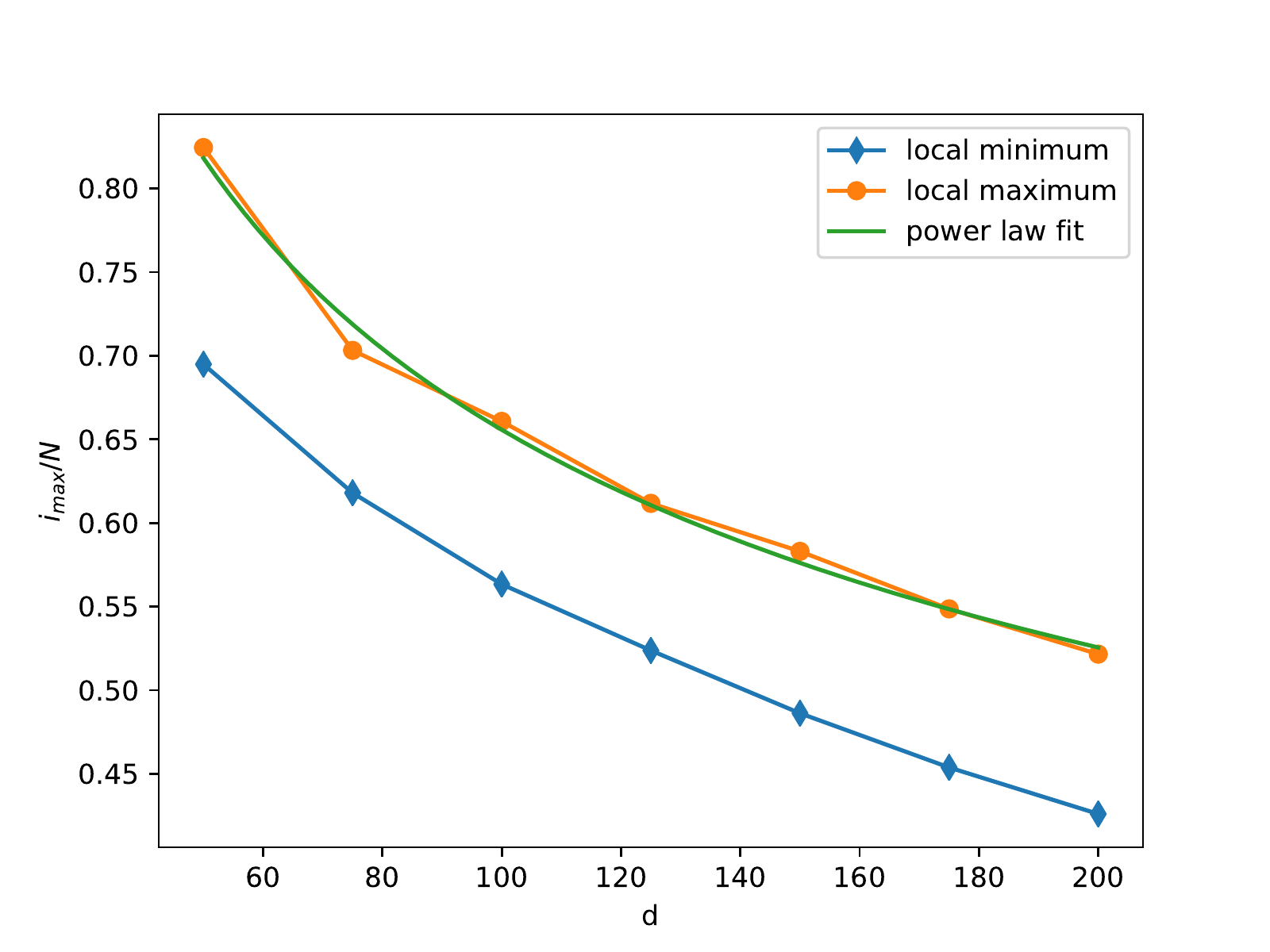}
	\caption{The local minimum and maximum of the condensation curve. The infectivity ratio is set to be $f=4$, all other parameters are chosen as in Fig. \ref{fig:close_wind}. Shown are the local minimum and maximum of the curve of the infected ratio in HS 2. The power-law fit is performed with a small exponent, $d^{-0.32}$, hence the correlation effects decay slower than the distance.}
	\label{fig:dist}
\end{figure}
 {Fig. \ref{fig:dist} shows that} both quantities decay according to an inverse power law $d^{-\alpha}$, with $\alpha \sim 1/3$, indicating that
 the correlation between the two hotspots decays much more slowly than their inverse distance. 
 To assess the effect of the diffusivity, we computed the condensation curve -- similar to Fig. \ref{fig:close_wind} 
 with fixed $f=4$, for different values of the diffusion constants $D$. 
 \begin{figure}[t]
 	\centering
 	\includegraphics[width=\linewidth]{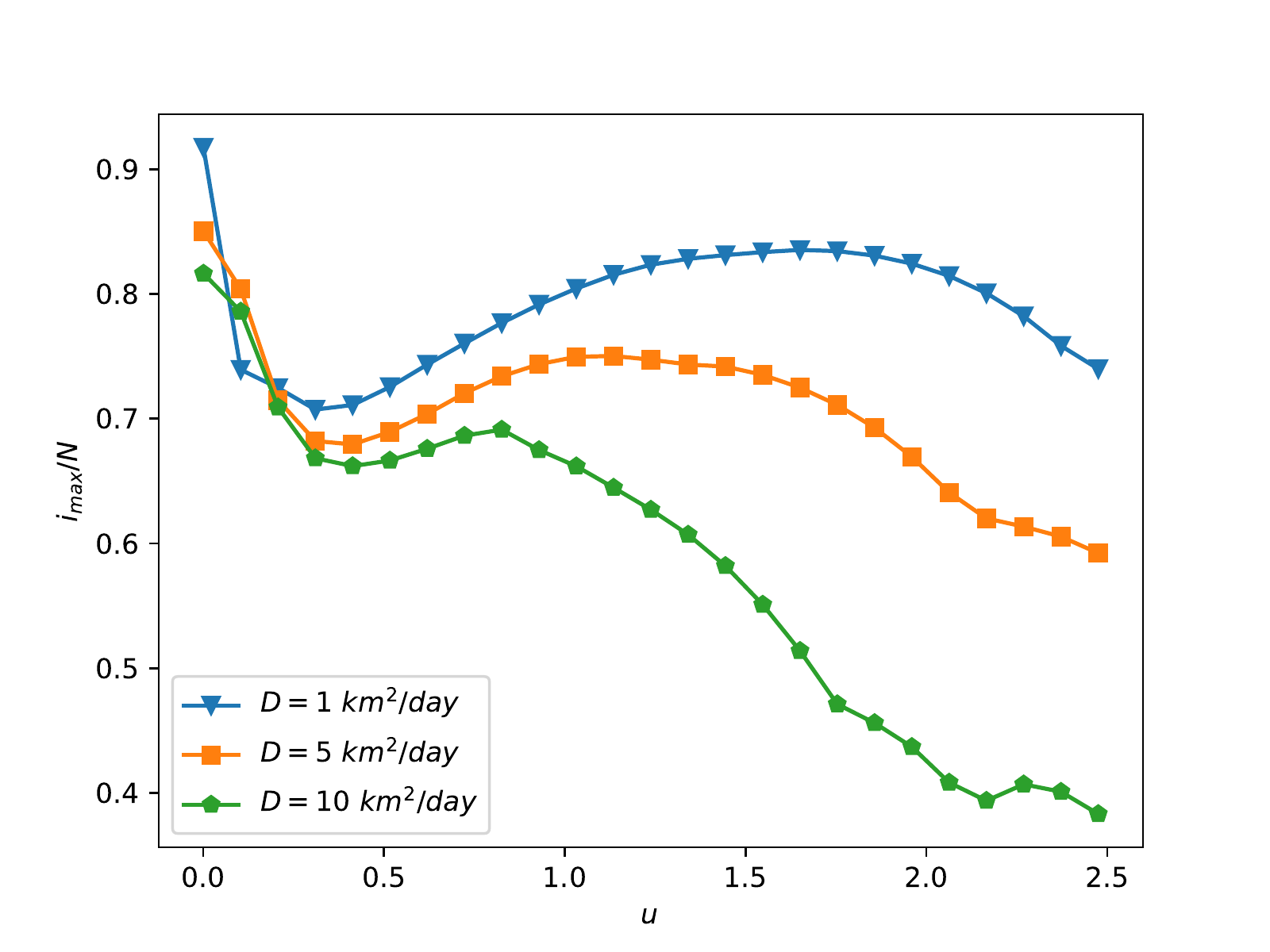}
	\caption{Effect of diffusion. We change the diffusivity and inspect its effect on the condensation curve for an intermediate high infectivity ratio $f=4$. For small $D$, we observe no qualitative change of the curve. $D=5\,km^2/day$ is the value we used throughout the rest of this paper. 
	When the Fischer speed $U_f=\sqrt{Df_1 \beta_0}$ approaches the reference wind speed $U_r$, we expect additional 
	interference effects which may lead a revival of infectivity in HS2. A detailed study of these phenomena is left to a future study.} 
	\label{fig:diff}	
\end{figure}
In Fig. \ref{fig:diff} we observe a quantitative effect of the diffusion parameter on the condensation curve, due to the fact that
diffusion smears out sharp spatial changes in population number, such as those observed at the hot-spot boundaries. 

Hence, the effect of increasing the diffusivity $D$ is similar to lowering the infectivity ratio $f$, as long as 
the diffusivity remains sufficiently small enough, meaning by this that the Fisher speed $U_f=\sqrt{D f_1\beta_0}$ 
remains well below the reference wind speed.  In the simulations carried out here, $U_f$ is always significantly 
smaller than $U_r$, hence these effects do not play any role. 
Whenever the condition $U_f \ll U_r$ is violated, non trivial interference
effects are expected, which may eventually lead to a revival of infectivity in the second hotspot. 
A detailed analysis of these effects warrants a separate study on its own, hence it is deferred to future investigations.
 %Even though a more systematic investigation is warranted, such "non-local" decay of spatial correlation 
 %is likely to represent a robust feature of epidemic condensation.   

% ------------------------------------------

% -----------------------------------------

%which are present for comparatively high $u$ - will decay when keeping $f$ and $u$ constant.
%However, surprisingly also the local minimum changes: The depth of the minimum becomes larger and the curve for small $u<1$ becomes steeper. So somehow we observe a stronger effect for larger distances which is counter-intuitive. A possible explanation could be the following: In this regime the decrease of the fraction of infected due to the emerging transparency at rising speeds $u>u_c$ is the dominant effect. \\
%At small differences the effective width of the hotspot might be larger than the true width $w=10$ because they are not independent. Therefore the transparency effect only gets strong at a higher $u$ and the minimum is not very deep. However, if the distance increases, the hotspots become more and more independent and the effective width is identical to the bare width $w=10$. Therefore the critical velocity becomes smaller, the transparency is present even for smaller $u$ and the minimum becomes deeper and the curve steeper. 

\section{Conclusions} Summarizing, we have evidenced a non-monotonic relation between
the wind speed and the peak intensity on the downstream hotspot as a function of the
infectivity ratio with respect to the upstream one. 
Despite its drastic simplification of the mechanism of human mobility, it is hoped that 
the non-monotonic "constitutive relation" revealed by the present ADSIR model, as
shown in Figures 4 and 5, may offer useful qualitative clues on the effects of spatial 
interference between infected hotspots.

\section*{Acknowledgements} SS kindly acknowledges funding from the European Research Council under the European
Union’s Horizon 2020 Framework Programme (No. FP/2014-2020)/ERC Grant Agreement No. 739964 (COPMAT). JD was supported by the ERASMUS program and Physics Advanced program of the Elite Network Bavaria (University of Regensburg, Germany). One of the authors would like to acknowledge useful discussions with Prof E. Marinari and G. Parisi in an early stage of the project. 

\bibliographystyle{ws-ijmpc}
\bibliography{adsir.bib}

%He also acknowledges illuminating discussions with Giorgio Parisi and Alessandro Vespignani.\\\\

\end{document}